# What Stabilizes the Intermediate Structure of an Amorphous Alloy?

Julián R. Fernández[*] and Peter Harrowell

*School of Chemistry, University of Sydney, New South Wales, 2006, Australia*
*and*
[*]*Comisión Nacional de Energía Atómica, Av. Libertador 8250*
*Capital Federal, Buenos Aires, Argentina*

Abstract. We present the results of simulation studies of a model binary metal-metalloid alloy in which we characterize and explain the local coordination structure, the intermediate structure associated with the packing of these coordination polyhedra and the thermal stability of the various structural elements of this model amorphous solid.

## 1. INTRODUCTION

Stability demands structure. There must, after all, be some feature of the particle arrangement that distinguishes the stable configuration from all the others. This is as true for amorphous solids as crystalline ones. In this paper we describe our efforts to characterise and explain the local and intermediate structure associated with the low energy amorphous configurations of a model metal-metalloid alloy.

The literature on the structure of amorphous alloys is comprised of three basic themes. The most universal of these is that of dense hard sphere packing. Bernal's classic work on random close packing [1] of spheres provided an explicit structure, complete with interstices, with which to think about an amorphous solid. The first attempt to extend this picture to mixtures of spheres saw Polk [2] propose that solutes occupied the holes that Bernal had characterised. This asymmetric treatment of the solvent and solutes was challenged on the grounds that there is clear evidence that the choice of solute changed the solute coordination [3]. We shall return to consider this chemical order as the third of our three themes. Numerical simulations of the optimum packing of mixtures of hard spheres offers one useful way forward. There have been a number of such calculations[4,5].

In contrast to large scale numerical density optimisation, Egami and Waseda [6] adopted the strategy of identifying the limit of stability of substitutionally disordered crystals on the basis of packing effects. They proposed a simple relationship between the radius ratio and the maximum concentration of solute at which the crystal is stable. The relationship has been shown [5] to provide a quite reasonable estimate of the stability limit for hard sphere mixtures for radius ratios in the range of $0.6 < r_B/r_A < 3.0$.

Recently, Miracle [7] has upgraded the original Polk idea. Instead of starting with a dense random packing of solutes, Miracle builds in the chemical order missing in the original theory by considering a dense packing (crystalline fcc, in the case of ref. [7]) of the solute-centered coordination clusters. Like Polk, other components are considered to occupy the interstitials of this structure. The Miracle model is noteworthy as it one of the first packing models that considers the nature of organisation beyond the nearest neighbour shell.

The problem of matching the local optimum with the global optimum is the core of the second theme which is organised about the significance of the icosahedron and polytetrahedral structures as local energetic optimums and the geometric frustration that results in trying to densely fill space with these shapes. In 1952, Franck [8] noted that the icosahedron represented the lowest energy coordination of an atom. Descriptions of the extended structure resulting from the inability to close pack such polyhedra have taken a variety of forms. Hoare [9] looked at building clusters of increasing intricacy out to the radial distance at which strain effects began to dominate. Noting that the tetrahedra can fill a curved 3D space, Sadoc [10] considered an approximate to the extended poly-tetrahedral structure (which includes icosahedral coordination) based on the projection of the regular structure in curved space down on to 3D. In 1992 Dzugatov [11] constructed a potential with a non-monotonic tail that specifically stabilised the icosahedral coordination and destabilised the close packed arrangements. This model has proved to be a useful tool in studying the stability of structures in the bulk and in clusters where local polytetrahedral ordering runs up against the geometrical restrictions associated with space filling.

Finally, we consider the role of a kind of order not mentioned in either of the two previous approaches. Many glass forming alloys exhibit quite marked chemical ordering, characterised by a preferential association of unlike species. The metal-metalloid alloys typically exhibit such order. Gaskell [3] proposed starting from this chemical order and suggested that the glass structure was comprised of a random mixture of the two stable $A_3B$ crystal packing - $Ni_3P$ and $Fe_3C$. This proposal, which we shall test below, is based on the proposition that chemical order so reduces the structural options that the crystalline arrangements dominate all solid structure.

In this paper we describe the results of simulations of a model binary alloy, inspired by the Ni-P system. Our strategy is to build up a structural description of the amorphous state and the associated rationalisation of this structure based only on what can be directly confirmed in our calculations. We find clear chemical ordering, although with little obvious signs of intermediate crystalline organisation. We find that packing considerations about the solute (B) and the solvent (A) dominate local and intermediate organisation, respectively.

## 3. MODEL AND ALGORITHM

The Lennard-Jones (LJ) potential for a mixture has the form

$$V_{ij}(r) = 4\varepsilon_{ij}\left[\left(\frac{\sigma_{ij}}{r}\right)^{12} - \left(\frac{\sigma_{ij}}{r}\right)^{6}\right] \quad (1)$$

where the sub-indices $i$ and $j$ could take the values A or B. We truncate the potential at a distance $2.5\sigma_{ij}$ and shift the potential so that it equals zero at the cut-off. (Here we shall set the masses of both components equal to $m$.) We shall work in the following reduced units throughout this paper: the unit of length is $\sigma_{AA}$, the unit of energy $\varepsilon_{AA}$, and the unit of time $\tau = \sigma_{AA}(m/\varepsilon_{AA})^{1/2}$. We shall follow Kob and Andersen (KA) [12] and set $\sigma_{AB} = 0.8$, $\sigma_{BB} = 0.88$, $\varepsilon_{AB} = 1.5$ and $\varepsilon_{BB} = 0.5$. This mixture at a composition of $x_B = N_B/N = 0.2$ has been studied extensively as a model glass former. Previously, we have reported on the crystalline phase [13] of the KA model as well as of more general binary Lennard-Jones models [14].

Molecular dynamics simulations have been carried out at constant NPT using a Nosé-Poincaré-Andersen Hamiltonian and a generalised leapfrog algorithm [15]. All calculations were performed at zero pressure. Enthalpy minimizations were carried out using a conjugate gradient scheme which ensures a fixed pressure.

## 4. RESULTS

### 4.1 The Stable and Metastable Crystal Structures

To establish some intuition about the stable structures available to this particular model, we shall begin with the crystal structures of the KA mixture. We have completed a study of the lattice energies of a range of LJ mixtures [14]. In Figure 1 we plot the lattice energies per particle of a number of crystal structures as a function of $\sigma_{AB}$ for $A_3B$ mixtures (i.e. $x_B = 0.25$). We have also calculated the density of states of the local potential energy minima of the amorphous mixture (see ref. [16] for details).

As reported previously [13], the most stable crystal state of the KA mixture over the composition range $0.0 \leq x_B \leq 0.5$ consists of coexisting face centered cubic (fcc) of pure A and the CsCl structure with composition AB. In order of ascending lattice energies we have the $PuBr_3$ structure, coexisting $Pd_2Zr$ structure and fcc, the cementite $Fe_3C$ structure and then the $Ni_3P$ structure. For reference, the lowest 'lattice energy' per particle obtained for the amorphous state following a similar enthalpy minimization is -7.92, significantly higher than the lattice energy of any of the crystalline states identified.

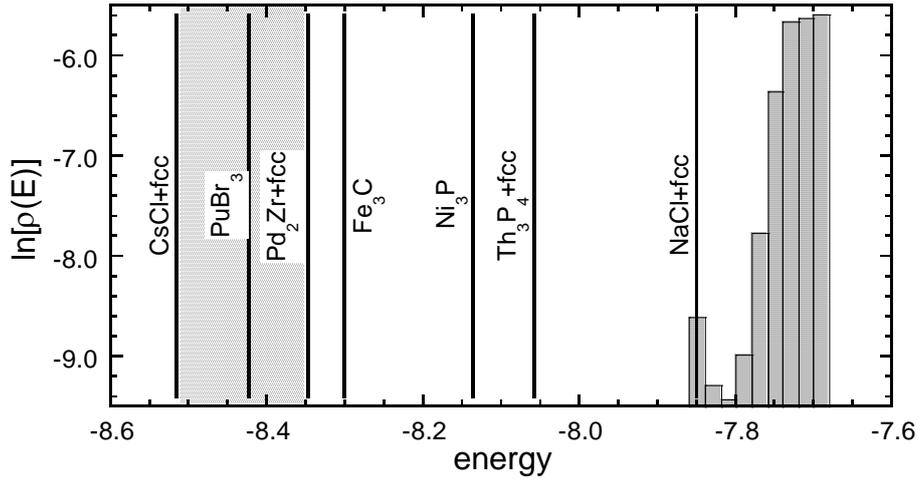

**FIGURE 1**. Lattice enthalpies per particle of a number of binary crystal structures for the KA model at a composition $x_B = 0.25$. The shaded region between "CsCl+fcc" and "$Pd_2Zr$+fcc" corresponds to a compositional continuum comprising of layers of fcc and CsCl structures. The shaded histogram on the right of the plot represents the density of states of the local amorphous minima.

In the CsCl structure each B particle lies in the center of a cube of eight A particles. In the $PuBr_3$, $Fe_3C$ and $Ni_3P$ structures, each B particle lies in the center of a tricapped trigonal prism (see Figure 2) consisting of nine A particles. As we shall see, the local structure in the various crystals is also found to dominate the local structure of the amorphous states.

We have estimated the energy of a finite sized two phase system consisting of pure A (fcc) and AB (CsCl) in which the incoherent interfacial energy is included [13]. We find that for $N \leq 1534$, the two crystal phases plus interface produce a higher energy than the amorphous state. Whatever kinetic impediments there are to crystallisation in the simulation, this result indicates that there can also be a significant reduction of the free energy difference between the amorphous and ordered states with decreasing system size.

## 4.2 Chemical Order and Intermediate Structure: Linking Coordination Polyhedra

In the KA mixture, the A-B interaction is strongly favoured over the B-B interaction. As a result, for $x_B < 0.20$ each B particle has only A nearest neighbours at low temperatures. After quenching the $A_3B$ mixture down to T = 0, we find (see Figure 2) a relatively small number of local coordination geometries about the B particles involving coordination numbers of 8 or 9.

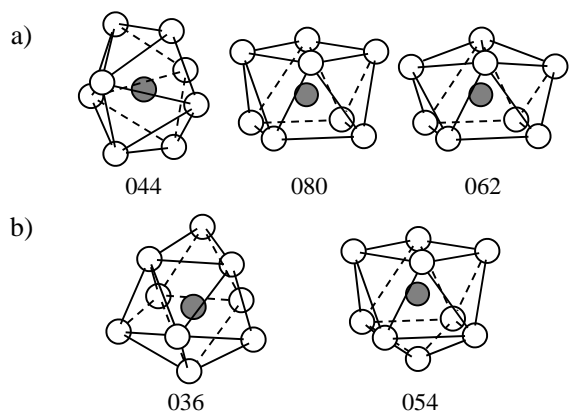

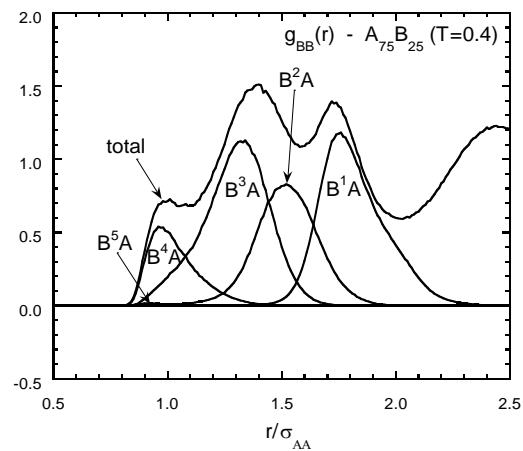

**FIGURE 2**. The most common a) 8-fold and b) 9-fold coordination geometries observed in the configurations of the local potential energy minima in the amorphous $A_3B$ mixture. The numerical code *nmp* means that the coordination polyhedra has $n$ x 3-fold, $m$ x 4-fold and $p$ x 5-fold vertices. Structure 036 is the tricapped trigonal prism (TTP) which is the form of solute coordination in the $Ni_3P$ and $Fe_3C$ crystal structures.

The coordination polyhedra are linked through the sharing of the A particles. To characterise these links we can define an $n$-fold 'bond' between B particles - $B^nA$ - when the two B particles share $n$ A neighbours.

As each B particle represents the centre of a polyhedron of A particles, then it follows that any two B particles that share four A neighbours represent two polyhedra sharing a 4-fold face. Similarly, two B particles that share three A neighbours correspond to adjacent polyhedra sharing a triangular face, and so on. We shall refer to such B particles pairs as "$B^nA$ bonds" where $n$ is 4, 3, 2 or 1 depending on the number of A neighbours shared by the two B particles.

Since each $B^nA$ bond can be interpreted in terms of the packing of adjacent polyhedra we have:

$B^1A$ = shared vertex

$B^2A$ = shared edge

$B^3A$ = shared face (triangular)

$B^4A$ = shared face (quadrilateral)

Unlike spheres, polyhedra generate multiple nearest neighbour lengths, depending upon whether the connection involves 1-, 2-, 3- or 4-fold bonds. This is immediately evident in the radial distribution function for B-B separations shown in Figure 3. The distinct lengths for the different types of $B^nA$ bonds show up quite clearly as distinct peaks.

**FIGURE 3.** The contributions to the B-B radial distribution function $g_{BB}(r)$ from the $B^nA$ bonds as described in the text with $n$ = 1-5 for the $x_B$ = 0.25 mixture at T = 0.4 . Note that the first peak is due almost exclusively to the $B^4A$ bonds and that the second peak is largely due to the $B^3A$ bonds which, at this composition, significantly exceed the $B^4A$ bonds in number.

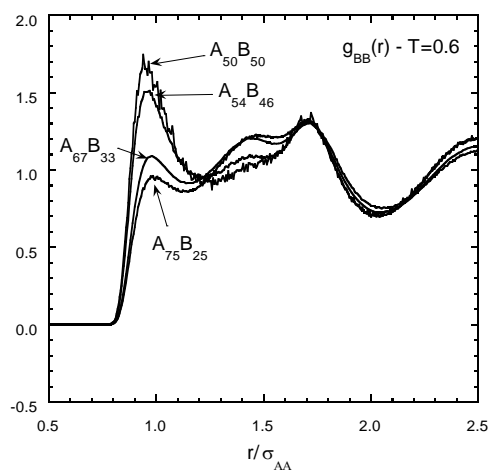

**FIGURE 4.** The B-B radial distribution function $g_{BB}(r)$ at T = 0.6 for the following compositions: $x_B$ = 0.25, 0.33, 0.46 and 0.5. Note the significant increase in the height of the first peak as $x_B$ is increased from 0.33 to 0.46 and the accompanying decrease in the height of the second peak.

The small first peak in the $A_3B$ mixture can thus be directly attributed to the relatively small number of polyhedra sharing square faces. We can also understand the anomalous temperature dependence of $g_{BB}(r)$ in which the height of the first peak decreases on cooling. If the triangulated coordination polyhedra are more stable at this composition, then we would expect the number of $B^4A$ bonds to decrease with the temperature.

As the number fraction $x_B$ of B species increases, each A particle is forced to accommodate more B neighbours into its first coordination shell. With each B particle trying to avoid the others, the geometrical possibilities for the A coordination quickly dwindle and one sees an increase in cubic structures and their associated shared square faces in Figure 4.

## 4.3 The Extended Organisation of Shared Triangular Faces

We would like to understand how the coordination polyhedra are packed together. To this end we shall look at the geometrical organisation of the $B^3A$ bonds corresponding to shared triangular faces. Our reasons for choosing the 3-fold bonds are i) they are more likely to be associated with rigid constraints than the shared vertex or edge, and ii) there is roughly 3 such bonds per solute (see Figure 5), a number that is neither too large to be difficult to visualised, nor too small to capture global organisation.

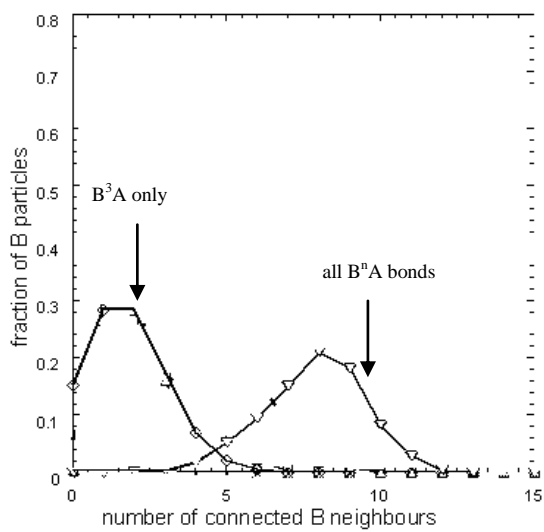

**FIGURE 5.** The distributions of the number of $B^nA$ (for all n) and $B^3A$ per B particle in the structures corresponding to local energy minima in the $A_3B$ mixture.

First we shall look at the spatial organisation of the 3-fold bonds in two $A_3B$ crystals - $Ni_3P$ and $Fe_3C$. These are shown in Figure 6a and b respectively. The $Ni_3P$ structure consists of 4 tricapped trigonal prisms in a ring of shared faces. These rings are then organised into parallel stacks. The lower energy $Fe_3C$ structure is made up of extended zig-zag strings of 3 fold bonds arising from stacks of single TTP's.

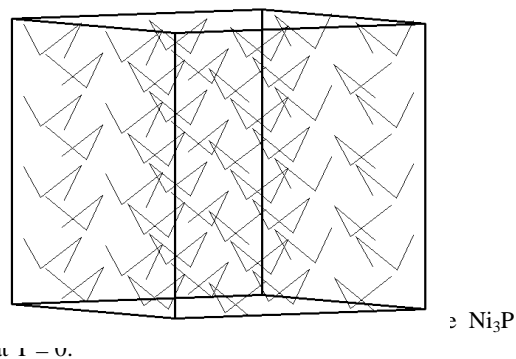

**Figure 6a.** The organisation of 3-fold bonds in the $Ni_3P$ crystal at $T = 0$.

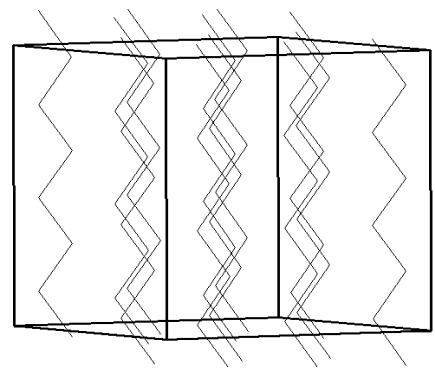

**Figure 6b**. The organisation of 3-fold bonds in the $Fe_3C$ crystal at $T = 0$.

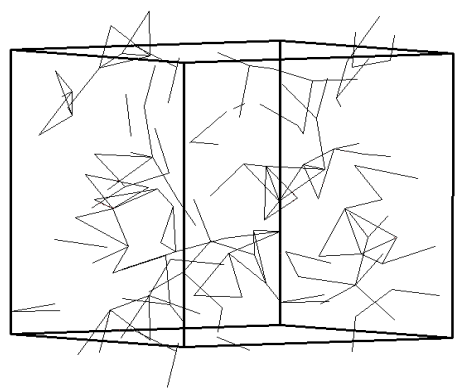

**Figure 6c**. The organisation of 3-fold bonds in an amorphous $A_3B$ alloy at $T = 0$.

A representative amorphous structure is also provided as Figure 6c. Despite the complexity, we can clearly see some distinctive features in the intermediate structure of the amorphous sample. There is a significant population of triangles of 3-fold bonds. It is quite common for two such triangles to share an edge and so produce a dihedral with an angle of roughly 120°. The extended strings of bonds appear to be far less constrained and exhibit a broad distribution of bond angles.

The distribution of angles between the bonds provides a quite distinctive measure of the intermediate structure associated with the organisation of the coordination polyhedra. In the cases of the two crystal structures, $Ni_3P$ and $Fe_3C$, there is only one bond angle for each, 68° and 113° respectively. In the case of the liquid $A_3B$ mixture we find that the bond angle distribution steadily sharpens on cooling, as shown in Figure 7.

suggestion [3] that the glass was made up of random mix of the two crystal arrangements of polyhedra. Instead we consider the shared A particles. There are, by definition, three shared A particles per $B^3A$ bond. When we consider the triplet of B's required to define the bond angle we can ask how many of the A's are shared between the coordination shells of all three B's? The possibilities are two, one or none. Each bond angle in our amorphous configurations can be categorised as having two, one or no A's common to all three B's. When we look at the angle distributions of these three classes of B triplets, as shown in Figure 8, we find each of our three peaks in the angular distribution arises from just one of these classes.

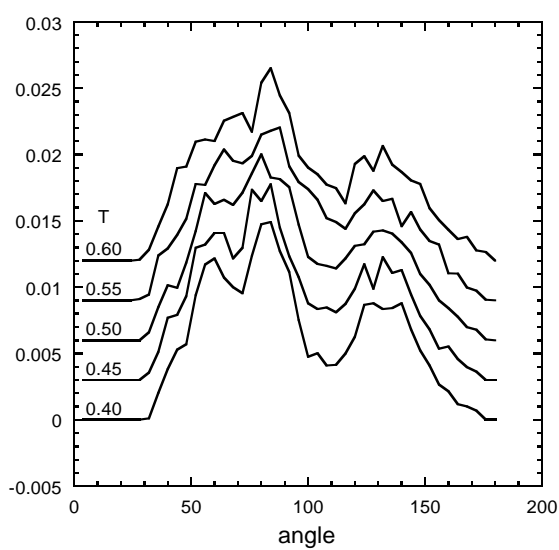

**Figure 7**. The distribution of angles between $B^3A$ bonds for the amorphous $A_3B$ mixture at temperatures of T = 0.60, 0.55, 0.50, 0.45 and 0.40.

In the limit of T = 0, arrived at by a conjugate gradient minimisation of the potential energy, we find the bond angle distribution to exhibit three distinct peaks (see Figure 8). There are two sharp peaks at 60° and 90° and a broader peak with a maximum around 135°. The presence of such distinct structural features associated with the intermediate structure of an amorphous alloy is noteworthy in itself. Clearly there is much to be learnt about alloy structures through the study of three body correlation functions, of which the angle distributions plotted in Figures 7 and 8 are one reduced version.

These sharp angle distributions suggest the presence of quite specific geometric structures. The systematic difference between the angle distributions in the amorphous state and those found in the $Ni_3P$ and $Fe_3C$ crystal structures do not support Gaskell's

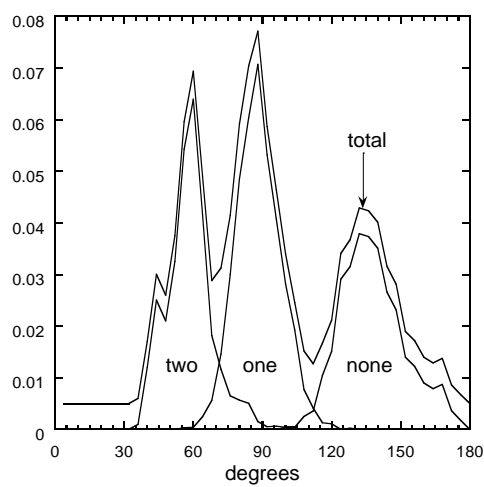

**Figure 8**. The distribution of angles between $B^3A$ bonds for the amorphous $A_3B$ mixture at T = 0 showing the contributions from triplets of B particles that share 2, 1 or 0 A particles.

The shared A particles impose quite strong geometric constraints on the arrangement of the B particles (and, hence, the arrangement of the coordination polyhedra about them). This can be seen in the sketches in Figures 9a and 9b of the cases in which two and one A particles, respectively, are shared between three B's. The two shared A particles sit above and below an equilateral triangle of 3-fold bonds between the B's. The origin of the sharp 60° peak becomes quite clear, as does the presence of triangles of bonds in the amorphous configuration shown in Figure 6c. A single shared A particle requires a large bond angle in order to accommodate the other non-shared A's. The absence of any shared A's represents the least constraint, accounting for the breadth of the angular distribution in this case.

Having spent so much of this paper focused on the geometry of the coordination of the B particles, it is important to emphasize where we have finally got to. Our explanation of the intermediate structure associated with the three body correlations between the B's rests on the consideration of the organisation of the A's and B's in the coordination shell of the shared A particles. In fact, all of the well defined intermediate structure we have found to date can be directly related to how the B particles arrange themselves about the A. Where there are only a few types of coordination about the B's (see Figure 2) there are a large number of possible A coordination structures due to the combination of compositional and structural possibilities. Understanding how an amorphous alloy, such as the one modeled in this paper, comes to 'select' the observed distribution of A-centered coordination structures is the core problem of intermediate structure stability in the chemically ordered glassy alloys.

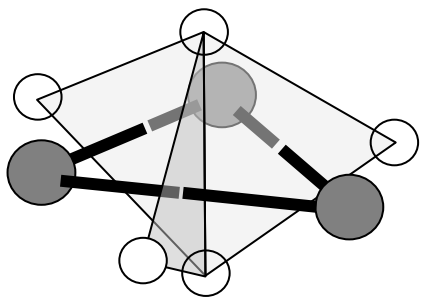

**Figure 9a**. A sketch of the arrangement of B (shaded circles) and A (open circles) particles when two A particles are shared between three B particles. The shaded triangles identify the three A particles shared between a pair of B particles to form a $B^3A$ bond.

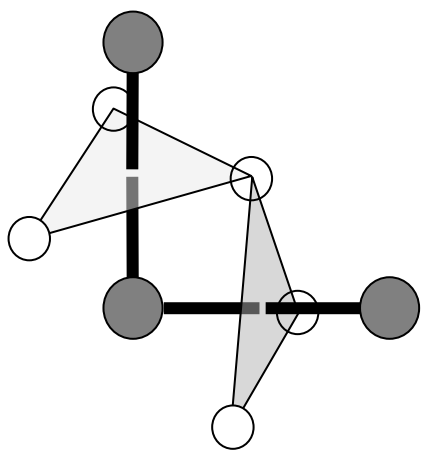

**Figure 9b**. A sketch of the arrangement of B (shaded circles) and A (open circles) particles when one A particle is shared between three B particles.

## 5. CONCLUSIONS

In this paper we have presented an account of the structure of the amorphous binary mixture introduced by Kob and Andersen. Our main results can be summarised as follows. At low temperatures we find almost all the B particles in either 8- or 9-fold coordination polyhedra. These polyhedra are largely triangular-faced with the result that the first peak of $g_{BB}(r)$, which reflects the number of shared square faces, is considerably smaller than the second peak. The proposal that it is the stability of these triangular-faced polyhedra that suppresses crystallisation of the CsCl phase gains support from the coincidence of the appearance of $B^4A$ bonds and the onset of crystallisation as the composition $x_B$ is increased.

Through the study of the angle distribution between 3-fold bonds linking the B particles we have shown that a sharp structural correlations exist, corresponding to well defined intermediate order generated by the packing of the coordination polyhedra. We have gone on to show how this feature of the intermediate structure is accounted for by the structure of the mixed coordination shells about the A particles. Much of the intermediate structure in the chemically ordered mixtures can be attributed to the distribution of B particles in the coordination shell of the A's. Understanding the distribution of compositions and structures among the A-centered clusters represents the outstanding problem with respect to developing a rational pictures of the intermediate order in these binary alloys.

## ACKNOWLEDGMENTS

We would like to thank Dan Miracle for valuable discussions and to acknowledge the support of the Australian Research Council and the Comisión Nacional de Energía Atómica of Argentina.